\documentclass[a4paper,aip,apl,amsmath,amssymb,reprint,superscriptaddress]{revtex4-1}
\usepackage{graphicx}
\usepackage{amsmath}
\usepackage{dcolumn}
\usepackage{bm}
\usepackage{bbold}
\usepackage{color}
\usepackage{amsfonts}
\usepackage{amssymb}
\usepackage{mathrsfs}
\usepackage{tabularx}
\usepackage{braket}
\usepackage{mathtools}
\usepackage{soul}			

\usepackage{caption}
\usepackage{subcaption}

\begin{document} 
\title{Inducing Strong Non-Linearities in a Phonon Trapping Quartz Bulk Acoustic Wave Resonator Coupled to a Superconducting Quantum Interference Device}
\author{Maxim Goryachev}
\affiliation{ARC Centre of Excellence for Engineered Quantum Systems, School of Physics, University of Western Australia, 35 Stirling Highway, Crawley WA 6009, Australia}

\author{Eugene N. Ivanov} 
\affiliation{ARC Centre of Excellence for Engineered Quantum Systems, School of Physics, University of Western Australia, 35 Stirling Highway, Crawley WA 6009, Australia}

\author{Serge Galliou}
\affiliation{FEMTO-ST Institute, CNRS, Univ. Bourgogne Franche Comte, ENSMM, 26 Chemin de l'\'{E}pitaphe, 25000, Besan\c{c}on, France.}

\author{Michael E. Tobar}
\email{michael.tobar@uwa.edu.au}
\affiliation{ARC Centre of Excellence for Engineered Quantum Systems, School of Physics, University of Western Australia, 35 Stirling Highway, Crawley WA 6009, Australia}

\date{\today}

\begin{abstract}A quartz Bulk Acoustic Wave resonator is designed to coherently trap phonons in a way that they are well confined and immune to suspension losses so they exhibit extremely high acoustic $Q$-factors at low temperature, with $Q\times f$ products of order $10^{18}$ Hz. In this work we couple such a resonator to a SQUID amplifier and investigate effects in the strong signal regime. Both parallel and series connection topologies of the system are investigated. The study reveals significant non-Duffing response that is associated with the nonlinear characteristics of Josephson junctions. The nonlinearity provides quasi-periodic structure of the spectrum in both incident power and frequency. The result gives an insight into the open loop behaviour of a future Cryogenic Quartz Oscillator in the strong signal regime.
\end{abstract}


\maketitle

\section{Introduction}

In the past ultra-stable low phase noise photonic frequency sources have been implemented for a range of applications including very high precision probes of fundamental physics, high precision oscillators as local oscillators for atomic clocks and advanced radar systems\cite{Nagel:2015dd,Abgrall:2016fq,Ivanov:2006yk,Ivanov:2009pv}. The best results at microwave and radio frequencies to date have been those derived from sapphire oscillators at room\cite{Ivanov:2006yk,Ivanov:2009pv} and cryogenic temperatures\cite{Nagel:2015dd,Abgrall:2016fq,locke} as well as lasers locked to optical cavities down converted through a frequency comb\cite{Millo,Fortier,Baynes,Diddams:10,Fortier:2011tt}. For example, such oscillators have been exploited to demonstrate possible violation of the Lorentz symmetry in the photon sector\cite{Nagel:2015dd}. Though the photonic devices have extraordinary stability, they can be used to test only certain range of physical phenomena, and devices where the resonance conditions depend upon mechanical motion of matter can test totally different sectors including frequencies of the bulk elastic waves, which are sensitive to the photon-, electron-, proton- and the neutron-sector. Moreover, recently new oscillator systems with rival performance of pure photonic systems have been demonstrated involving high-$Q$ frequency stable phonon systems. Such systems include but are not limited to, phonon lasers \cite{phononlaser}, optomechanical systems \cite{luan,seok,VahalaAPL} and phonon induced Brillouin scattering devices \cite{VahalaScience,Tomes,Grudinin,Lee,Li}.

Traditionally the most stable oscillators based on acoustic resonance have been realised from technology based on Bulk Acoustic Wave (BAW) resonators, which trap phonons in a cavity in a similar way to a Fabry-P{\'e}rot cavity. These techniques have been perfected for decades allowing precision room temperature oscillators and related devices, culminating in $Q\times f$ products as high $2\times10^{13}$~Hz and oscillator frequency stability reaching below $10^{-13}$ between 1 and 10 seconds of integration time\cite{Salzenstein:2010aa}. Only recently has this technology has been extended to cryogenic temperatures attaining $Q\times f$ products as high $2\times10^{18}$~Hz. Thus, such cryogenic BAW systems shows great potential for use in applications that require precision control, measurement, and sensing at the quantum limit. This is mainly due to the relatively high mechanical frequencies and extremely high $Q$-factors achievable at cryogenic temperatures ($Q \approx 10^{10}$)\cite{ScRep} for frequencies ranging from a MHz to tens of GHz, beyond the capability of any other competing technology compared in Aspelmeyer et al and others\cite{Aspelmeyer2014,Renninger,quartzPRL,Teufel:2011aa}. Such high-Q devices with nonlinear properties are very valuable in many areas of physics. This includes quantum metrological applications where nonlinearities are employed to perform quantum state preparation\cite{Hofheinz:2009aa}. In regards to creating low noise oscillators, nonlinear processes are used to create a limit cycle in feedback frequency to maintain stable output power.

The further improvement of BAW oscillators can only be achieved by cooling the resonators and reducing the resonator flicker phase self-noise, since this is the dominant noise source both at cryogenic and room temperature. The influence on the frequency stability of high-performance quartz oscillators on time scales of order $1-50$ sec is well-documented and it has been observed that the flicker self-noise decreases with decreasing power of the incident signal, and our recent results confirm that the resonators are thermal noise limited, and flicker-free without the carrier\cite{Goryachev:2014ab}. Thus, the noise in the quartz oscillator is dependent on power, with the white noise floor decreasing with power, while the flicker noise increases. The best quartz typically has frequency instabilities of better than $10^{-13}$ limited by flicker fluctuations. However at cryogenic temperatures the white noise floor is reduced by $40$~dB, allowing a much lower oscillator power and a large reduction of the flicker noise so the increased $Q$-factor may be exploited. Assuming the typical phase noise of $-130$~dBc/Hz at $1$~Hz Fourier frequency limited by the resonator self noise, the increase in $Q$-factor at 4K should see frequency instabilities as low as $2\times10^{-16}$\cite{PhysRevX.6.011018}, and if this can be pushed down further due to power optimisations it is strongly feasible to push the stability into the $10^{-17}$ regime. The other advantage at low temperatures, is the significantly reduced temperature coefficient, in the standard quartz BAW, the coefficient is annulled at around 1 K in operating temperature\cite{ours}. However, this value may be raised by varying the cut angle with respect to the anisotropy of quartz\cite{JBon}.

Although, in general all devices are nonlinear in nature, the feasibility to achieve nonlinear regimes are usually limited for a variety of reasons. For example, these regimes usually require considerably high amounts of incident power that may also cause nonlinear effects in auxiliary components, induce noise processes such as flicker noise, and cause the heating. So, in order to reduce this threshold of nonlinearity, it is required to reduce the system losses and increase nonlinear interactions. These requirement, in particular for BAW devices, usually contradict each other leading to trade offs. Another possible solutions is to utilise ultra high Quality factor resonators and couple them to a nonlinear superconducting circuit with a low threshold power. The former can thus provide very narrow spectral lines, while the latter adds the strong nonlinearities required at low powers. Thus, our goal to achieve better performance is to couple BAWs at low temperatures\cite{Goryachev:2014ab} to SQUID amplifiers\cite{SQUIDbook} and operate them with low power oscillations of order $-80$ dBm. To further increase the power output, a chain of regular cryogenic ultra-low noise amplifiers will be used allowing us to generate reasonable power of order $10$ dBm, with fractional frequency instabilities of $10^{-16}$ or better. In this work we make the first step towards this goal and show the large signal characteristics of a SQUID coupled to a BAW resonator. Similar SQUID-mechanical resonator systems have been used in the past as Gravity Wave detectors\cite{Xu:1985aa,Pleikies:2007aa,Goryachev:2014ac} and more recently as quantum hybrid systems\cite{OConnell:2010fk,Woolley:2016aa} but only in the non-driven or weakly driven regimes.

\section{Results: Towards a Cryogenic Quartz Oscillator}

One of the major requirements for feedback oscillators is an existence of a nonlinear element used to to create a limit cycle by transferring the main tone energy to higher harmonics. Despite the long history of classical feedback oscillators, the latter problem has recently emerged again for designing of CQO. Typically a CQO operating around $7$K has an extremely high power dependence with a power coefficient of frequency of order $1$ Hz$\mu$W at $-30$~dBm of input power, about 1000 times higher than a room temperature oscillator\cite{eugene}. A solution is to utilise an ultra high Quality factor resonator coupled to a nonlinear superconducting circuit to lower the threshold power. The BAW resonator can thus provide very narrow spectral lines, while the superconducting circuit adds the strong nonlinearities required at low powers. In this work we utilizes a Superconducting Quantum Interference Device (SQUID) circuit\cite{SQUIDbook} coupled to a high $Q$ BAW Cavity\cite{Goryachev:2014ab}.

\begin{figure*}
     \begin{center}
            \includegraphics[width=1.0\textwidth]{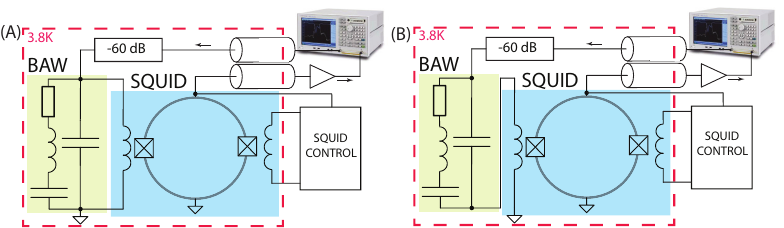}
            \end{center}
    \caption{ Experimental setups representing two ways to connect a BAW resonator and a SQUID amplifier: (A) parallel connection, (B) series connection.}
       \label{setupSIG}
\end{figure*}

The connection topologies involve a SQUID input coil, a two electrode BAW resonator and an external signal source. The parallel connection is realised when all three devices share the same voltage (see Fig.~\ref{setupSIG} (A)), and the series connection, when the three devices form a single current contour (see Fig.~\ref{setupSIG} (B)). These topologies may be understood as loaded feedback QCO with an open-loop. In both cases, the SQUID-BAW system is cooled down to $3.8$K with a conventional pulse-tube cryocooler. The acoustic resonator is an SC-cut\cite{pz:1988zr} BVA (electrodeless) \cite{besson2} quartz BAW device {\color{black}(1 mm thick, 30 mm diameter)}, whereas the SQUID is a commercial Niobium amplifier {\color{black} for which the current bias is applied across the loop. The SQUID amplifier does not have additional noise cancellation, and its flux bias is static. Bias parameters are tuned to maximise gain in the small signal case that is estimated to be $1.2$~MOhm in terms of transimpedance relating input current and the output voltage. } 
Both devices have been used for Nyquist noise measurements at liquid helium temperatures\cite{Goryachev:2014ab}. The signal is fed through a long coaxial line ending at a $-60$~dB cold attenuator. The output signal of the cold part of the DC SQUID amplifier is retrieved via a micro-coaxial line, which is connected to and read out by a room temperature amplifier. All the data is acquired by a Vector Network Analyser locked to a Hydrogen maser providing extra frequency stability over long averaging times. Long averaging times are required to keep the measurement bandwidth as low as $3-10$~Hz in order to avoid system ringing due to very high Quality factors.

The BAW resonator equivalent model may be represented by an equivalent circuit comprising a number of motional branches and a shunt capacitance. Each motional branch is a series connection of resistive, capacitive and inductive components. In this work we limit the investigation to a few low order modes that fall into the frequency range of the SQUID amplifier. The nonlinear response of mechanical resonators can be usually approximated by the Duffing model\cite{Gagnepain,nonlin,Tiers2,nosek} arising from the nonlinear elastic terms of the constitutive equations or thermoelectroelasticity\cite{Tiersten1}. This holds true for the ultra-high Quality factor cryogenic BAW resonators under investigation in this work\cite{SunFr,quartzPRL}, with the exception of resonators which are not swept of impurities and as a result have a large amount\cite{quartzJAP}.

\begin{figure*}
     \begin{center}
            \includegraphics[width=1.0\textwidth]{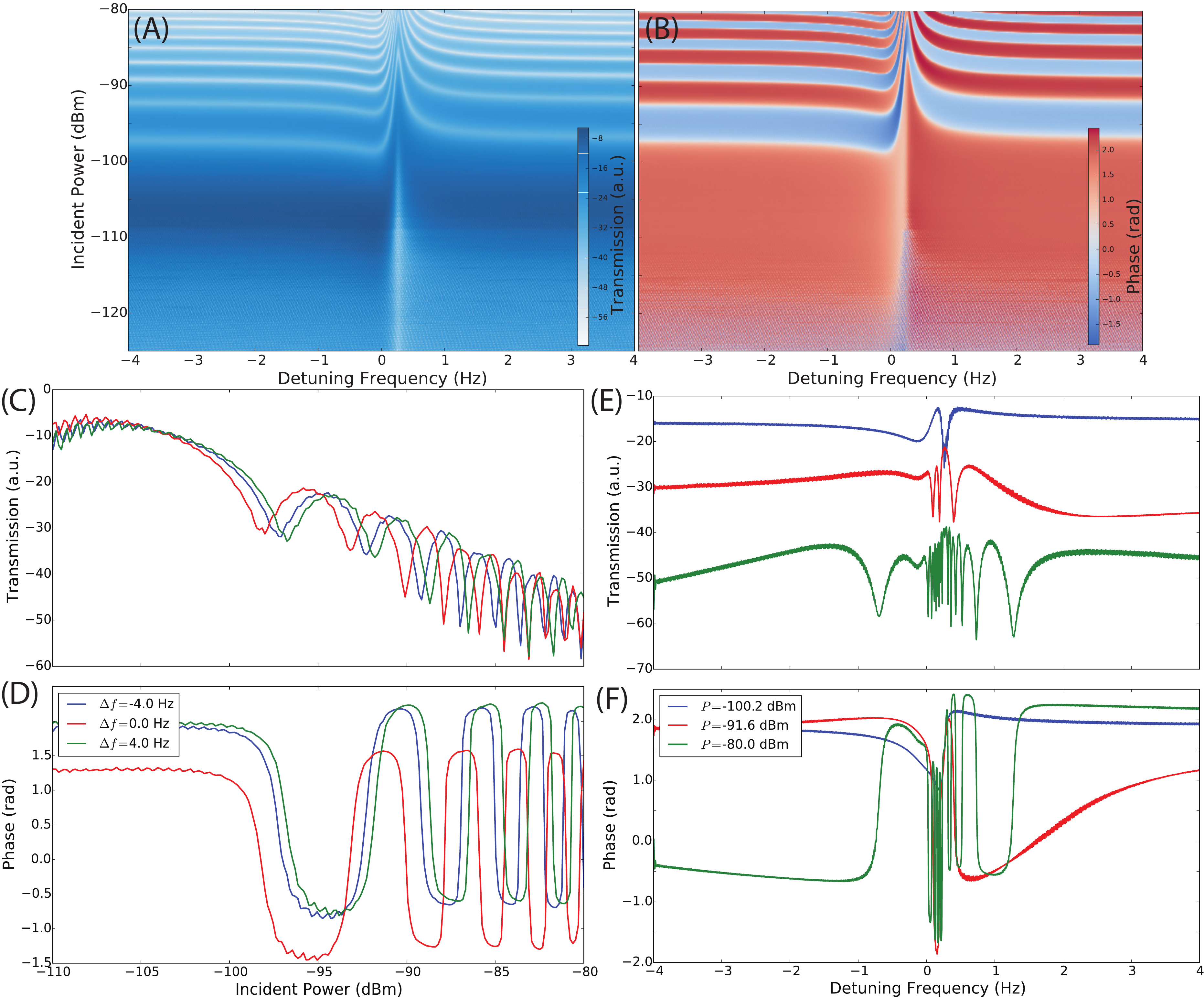}
            \end{center}
    \caption{Top: Amplitude (A) and Phase (B) response of the parallel system as a function of the incident power in the vicinity of the 3rd overtone of the C bulk acoustic mode ($4.993027$~MHz). Bottom Left: Amplitude (C) and Phase (D) response as a function of power for three specific values of the detuning frequency from the data represented on the left. Bottom Right: Amplitude (E) and Phase (F) response for three specific values of the incident power from the data represented on the left.}
   \label{fig2}
\end{figure*}

The amplifier is a standard DC Niobium SQUID device with an nominal input inductance of $400$~nH and the transfer coefficient of about $300\mu$V$/\phi_0$. The experiment has been repeated with two identical amplifiers with similar results. For each experiment the SQUID bias has been kept in such a way that the system works on the linear part of its characteristic. 

\begin{figure*}
     \begin{center}
            \includegraphics[width=1.0\textwidth]{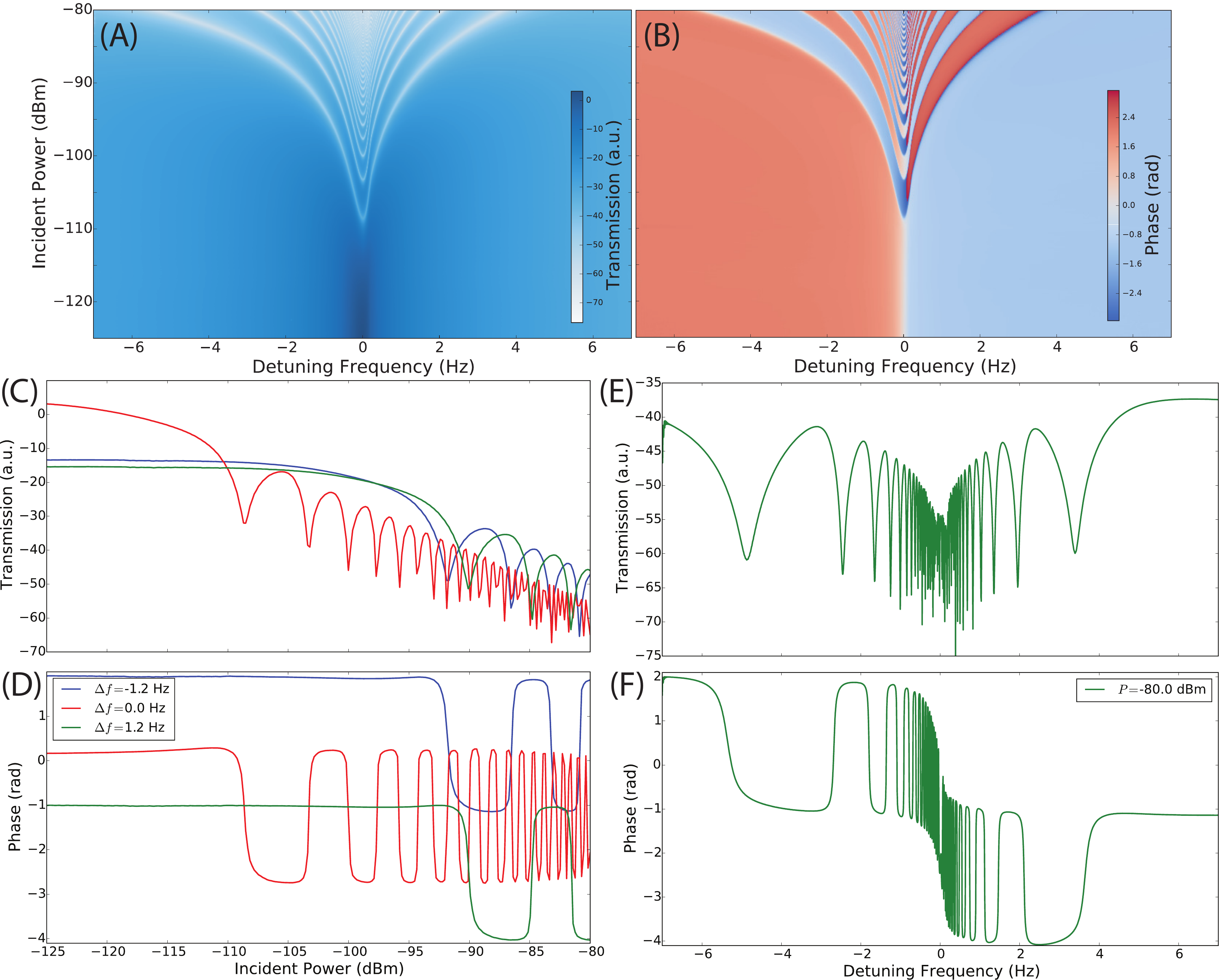}
            \end{center}
    \caption{Top: Amplitude (A) and Phase (B) response  of the series system as a function of the incident power in the vicinity of the 5th overtones of the A (quasi-longitudinal) mode ($15.732444$~MHz). Bottom Left: Amplitude (C) and Phase (D) response as a function of power for three specific values of the detuning frequency from the data represented on the left. Bottom Right: Amplitude (E) and Phase (F) response for three specific values of the incident power from the data represented on the left.}
   \label{fig3}
\end{figure*}

Fig.~\ref{fig2} shows the amplitude and phase response of the parallel system as a function of the incident power and frequency of excitation in the vicinity of the 3rd overtone of the C bulk acoustic mode ($4.993027$~MHz), with the periodicity of the power dependence apparent and demonstrated further for three values of the detuning signal. These dependencies can not be described by the Duffing model used to describe the response of bare mechanical resonators and requires full trigonometric function representation. It is seen that in both frequency and power the system response is best described by chirp functions. The applied powers are too low to induce the resonator own nonlinearity, and the same type of response is observed for all low frequency BAW modes with the nonlinearity becoming apparent at different power levels depending on the mode Quality factor and motional resistance (which describes the electromechanical coupling). {\color{black} For the 3rd overtone of the C bulk acoustic mode, the Quality factor is $4.9\times 10^7$ as found from Nyqvist noise spectrum\cite{Goryachev:2014ab}.}

Fig.~\ref{fig3} shows the amplitude and phase response of the series system as a function of the incident power and frequency of excitation in the vicinity of the 5th overtone of the A (quasi-longitudinal) mode ($15.732444$~MHz). In addition to quasi-periodic structure of the spectrum, the system demonstrates symmetry around the resonance frequency in the linear regime: mirror symmetry for the magnitude and diagonal symmetry for the phase. Also, similar to the parallel connection results, the intrinsic BAW nonlinearity is not observed. {\color{black} For this mode, the Quality factor is $27\times 10^7$ obtained from Nyqvist noise spectrum\cite{Goryachev:2014ab}.}

The system Hamiltonian (in the units with $\hbar=1$) can be written based on the equations of motion: 
\begin{multline}
	\label{FF0001G}
		H = \omega_m a^\dagger a + \omega_J\sum_i\Big(\frac{q^2_i}{2}+\frac{\delta^2_i}{2}-\xi\cos\delta_i\Big)-\omega_J\delta_1\delta_2\\
 -\omega_J(\phi_+\delta_1+\phi_-\delta_2) - g(a^\dagger + a) (\delta_1-\delta_2),
\end{multline}
where $a$ ($a^\dagger$) is an annihilation (creation) operator for an acoustic mode, $\omega_m$ is an angular frequency of the mechanical mode, $q_i$ and $\phi_i$ are conjugate variables for two SQUID branches, $\omega_J=1/\sqrt{C_JL_J}$ is the Josephson junction plasma frequency, $\phi_+$ and $\phi_-$ are properly scaled biasing current and flux, $\xi = \frac{I_0}{\phi_0}L_J$. The charge terms containing $q_i$ may be removed since the effect of shunting capacitances is negligible at the working frequencies. {\color{black} It is also worth noting that none of the higher order modes of the resonator is an integer or rational multiple of the modes of interest. The deviation from being integer multiples of the fundamental mode for quartz resonators is due to the piezoelectric effect.} Nevertheless, direct simulation of this system is associated with certain difficulties.
For example, low dimensional nonlinear physical models are often treated asymptotically using perturbation techniques, which utilize only one higher harmonic\cite{Krylov,Nyfeh}. Though this technique is able to give adequate approximations for weakly nonlinear systems, in contrast systems with strong nonlinearities require complex numerical techniques such as the Harmonic Balance approach. The Harmonic Balance approach is a powerful technique to simulate the steady-state response of nonlinear systems in the frequency domain\cite{Nakhla:1976aa,Gilmore:1991aa}. Generally, this numerical approach splits a system into a linear and nonlinear parts and requires representation of each variable in a series of harmonics. Whereas the linear part represents the dynamical response at each frequency, the nonlinear part mixes the harmonic components. As a result the system is represented by a set of nonlinear algebraic equations that are solved numerically. The more advanced versions of this methods have been implemented for solving nonlinear circuit problems\cite{agilent1,MO1}. These types of software for designing nonlinear electrical circuits may be used for further numerical system analysis and design of future related devices.

\section{Discussion}

Understanding of the nonlinearity of a SQUID amplifier coupled to a high-$Q$ resonator is important for designing of a future Cryogenic Quartz Oscillators. The experiments presented in this work present results for open loop response of possible CQO topologies. Being a necessary condition for any feedback oscillator to create a limit cycle, the degree of nonlinearity controls the phase noise budget trade off. The strong low power nonlinearity keeps the oscillator circulating power low, which will typically reducing the flicker noise (the main limit in room temperature oscillators) but increasing the effect of thermal fluctuations due to the weaker signal to noise ratio of the acoustic resonance frequency determination. Weaker nonlinearity needs higher circulating power for saturation, thus minimising the thermal noise but inducing excess flicker noise\cite{Gagnepain:1983aa}. For CQO, thermal noise is naturally reduced by orders of magnitude by keeping the temperature low, so the SQUID-BAW system gives a way to reduce the oscillator power significantly to keep flicker noise as low as possible. Thus, this work in an enabling step towards producing high frequency stability\cite{Goryachev:2013ly} beyond the limit achieved at room temperature devices\cite{Salzenstein:2010aa} required for some test of fundamental physics e.g. the Lorentz invariance in the neutron sector\cite{PhysRevX.6.011018}.
 
This work was supported by the Australian Research Council Grant Numbers CE170100009 and DP160100253. SG would like to thank ENSMM for its financial support.

\hspace{10pt}

%

\end{document}